\documentclass[preprint]{elsarticle}
\usepackage{graphicx}
\usepackage{amsmath}
\usepackage{amsfonts}
\usepackage{hyperref}
\usepackage{color}
\usepackage{booktabs}

\allowdisplaybreaks[1]

\begin{document}

\begin{frontmatter}

\title{Numerical model\\ for macroscopic quantum superpositions \\based on phase-covariant quantum cloning}

\author{A. Buraczewski\fnref{elka}}

\author{M. Stobi\'nska\fnref{mpl,itp2}\corref{ms}}
\ead{magda.stobinska@mpl.mpg.de}

\fntext[elka]{Faculty of Electronics and Information Technology,
  Warsaw University of Technology, Warsaw, Poland}

\fntext[mpl]{Max Planck Institute for the Science of Light, Erlangen, Germany}

\fntext[itp2]{Institute for Theoretical Physics II,
  Erlangen-N\"urnberg University, Erlangen, Germany}

\cortext[ms]{Corresponding author: Magdalena Stobi\'nska}

\begin{abstract}

Macroscopically populated quantum superpositions pose a question to what extent macroscopic world obeys quantum mechanical laws. Recently such superpositions for light, generated by optimal quantum cloner, were demonstrated. They are of fundamental and technological interest. We present numerical methods useful for modeling of these states. Their properties are governed by Gaussian hypergeometric function, which cannot be reduced to neither elementary nor easily tractable functions. We discuss the method of efficient computation of this function for half integer parameters and moderate value of its argument. We show how to dynamically estimate a cutoff for infinite sums involving this function performed over its parameters. Our algorithm exceeds double precision and is parallelizable. Depending on the experimental parameters it chooses one of the several ways of summation to achieve the best efficiency. Methods presented here can be adjusted for analysis of similar experimental schemes.

\end{abstract}


\end{frontmatter}

\section{Program summary}

\begin{list}{}{\setlength{\leftmargin}{0pt}}
\item\emph{Program title:} MQSVIS
\item\emph{Catalogue identifier:} \verb|AEMR_v1_0|
\item\emph{Program summary URL:} \url{http://cpc.cs.qub.ac.uk/summaries/AEMR_v1_0.html}
\item\emph{Program obtainable from:} CPC Program Library, Queen's University, Belfast, N. Ireland
\item\emph{Licensing provisions:} Standard CPC licence, \url{http://cpc.cs.qub.ac.uk/licence/licence.html}
\item\emph{No. of lines in distributed program, including test data, etc.:} 1643
\item\emph{No. of bytes in distributed program, including test data, etc.:} 13212
\item\emph{Distribution format:} tar.gz
\item\emph{Programming language:} C with OpenMP extensions (main numerical program), Python (helper scripts)
\item\emph{Computer:} modern PC (tested on AMD and Intel processors), HP BL2x220
\item\emph{Operating system:} Unix/Linux
\item\emph{Has the code been vectorized or parallelized?:} yes (OpenMP)
\item\emph{RAM:} 200 MB for single run for $1000\times 1000$ tile
\item\emph{Running time:} 1-2h for $1000\times 1000$ tile, depending on the values of parameters
\item\emph{Classification:} 4.15, 18
\item\emph{Nature of problem:} Recently macroscopically populated quantum superpositions for light, generated by optimal quantum cloner, were demonstrated. They are of fundamental and technological interest. Their properties are governed by Gaussian hypergeometric function ${}_2F_1$ of half-integer parameters, which cannot be reduced to neither elementary nor easily tractable functions. Computation of photon number distribution, visibility and mean number of photons, necessary for characterization of these states, requires evaluation of infinite sums involving this function performed over its parameters.
\item\emph{Solution method:} MQSVIS program suite computes various quantum indicators, such as photon number distribution, visibility, mean number of photons, variance for macroscopic quantum superpositions of light. It takes losses (modeled with a beamsplitter) and imperfect photodetection (modeled with a Weierstrass transform applied to the photon number distribution) into account. Cutoffs of the infinite hypegeometric sums are estimated dynamically, and precision is enhanced with computation of expressions in logarithmic form. Depending on the experimental parameters, program chooses one of the several ways of summation to achieve the best efficiency. Program is parallelized using OpenMP standard, which ensures best utilization of multicore processors, and splits the work into tiles computed with different nodes of a computer cluster. This allows to compute the required indicators for realistic values of the parameters.
\end{list}

\section{Introduction}

Macroscopic quantum superposition (MQS) and entanglement were pointed out by E. Schr\"odinger in 1935. He posed a \textit{gedanken experiment} where the quantum formalism was applied to a macroscopic object -- a cat and a microscopic object -- a two-level radioactive atom, describing them in a joint quantum superposition \cite{Schroedinger1935}. These objects were closed in a box and if the atom decayed at a random time, additional mechanism installed inside was activated to kill the cat. The conclusion was striking: the cat was dead and alive at the same time, as long as no one opened the box. It revealed the phenomenon of quantum entanglement: the state of the cat is random (dead or alive) but perfectly correlated with the state of the atom (decayed or not). It also planted a question to what extent macroscopic objects obey laws of quantum mechanics.  

Small MQS were produced in: nanoscale magnets, laser-cooled trapped ions, photons in a microwave cavity, $C_{60}$ molecules, superconducting devices \cite{Friedman2000} and macroscopic-size diamonds cooled \cite{Gisin2011} and in room temperature (2011) \cite{Lee2011}. In 2007, macroscopically populated polarization entangled superpositions of light were generated in the room conditions by optimal quantum cloning (OQC). It is based on parametric frequency down conversion (PDC) \cite{Nagali2007,DeMartini-PRL} where a higher energetic blue photon in a given polarization is turned, with some probability, into two lower energetic red photons by a nonlinear crystal preserving the polarization. The more the crystal is pumped with the blue photon beam, the higher the probability of conversion, resulting in multi-photon output. OQC does not violate the no-cloning theorem \cite{Wootters1982}, which states that an unknown quantum state (here: polarization) cannot be copied perfectly. It makes imperfect copies, characterized by a cloning fidelity lower than one \cite{Scarani2005}. There are two kinds of MQS of light produced by OQC: the micro-macro singlet state, where a single photon is entangled with a ``macroscopic qubit'', and the bright entangled squeezed vacuum \cite{DeMartini-PRL,Masha}. They contain $10^5$-$10^{13}$ photons.

MQS are promising alternative for quantum technologies. They allow exploring the quantum-to-classical transition \cite{Zurek2003,Vitelli2010}, efficient interaction with matter and photons \cite{Jedrkiewicz2011}, studying the principle of quantum measurement and form a non-Gaussian class of quantum states, necessary for fault tolerant quantum computing \cite{Bartlett2002}. Importantly, they give hope for a loophole-free Bell inequality test \cite{Stobinska2011} enabling the ultimate test of the quantum theory. 

Theoretical and experimental analysis of MQS is challenging due to their large complexity and fast decoherence \cite{Bartlett2002}, both scaling exponentially with Hilbert space dimension. Commercial numerical tools (Matlab, Wolfram Mathematica) and libraries (BLAS, NetLib, LAPACK) model MQS only for small populations. Intuitions based on these results are often misleading for high populations \cite{DeMartini-PRL,Sekatski2009}. Here, slowly convergent multiple infinite hypergeometric series arises, which is intractable for these applications. This results in lack of commonly accepted model of MQS. 

In this paper, we develop numerical model for MQS of light and discuss the method of efficient computation of Gaussian hypergeometric function for half-integer values of its parameters and moderate value of the argument. We show how to dynamically estimate a cutoff for infinite sums involving this function performed over its parameters. Our algorithm offers precision exceeding double precision by several orders of magnitude. Depending on the experimental parameters it chooses one of the several ways of summation to achieve the best efficiency. It is parallelizable, which allows better utilization of multi-processor computers. We tested our numerical results with recently developed analytical model for MQS for certain regimes of parameter values \cite{Stobinska2011}. The numerical methods presented here can be adjusted for analysis of similar experimental schemes. 

This paper is organized as follows. In Section~\ref{macro} we introduce MQS, their properties and show the origins of hypergeometric function.
In Section~\ref{sec:numerical_methods} we discuss methods used in numerical model of MQS. Section~\ref{sec:examples} includes examples: computation of MQS photon number distribution and distinguishability.

\section{Theoretical background}\label{macro}

\subsection{Entangled Macroscopic Quantum Superpositions}

\begin{figure}
  \begin{center}
    \raisebox{2.5cm}{a)}
    \includegraphics[height=3cm]{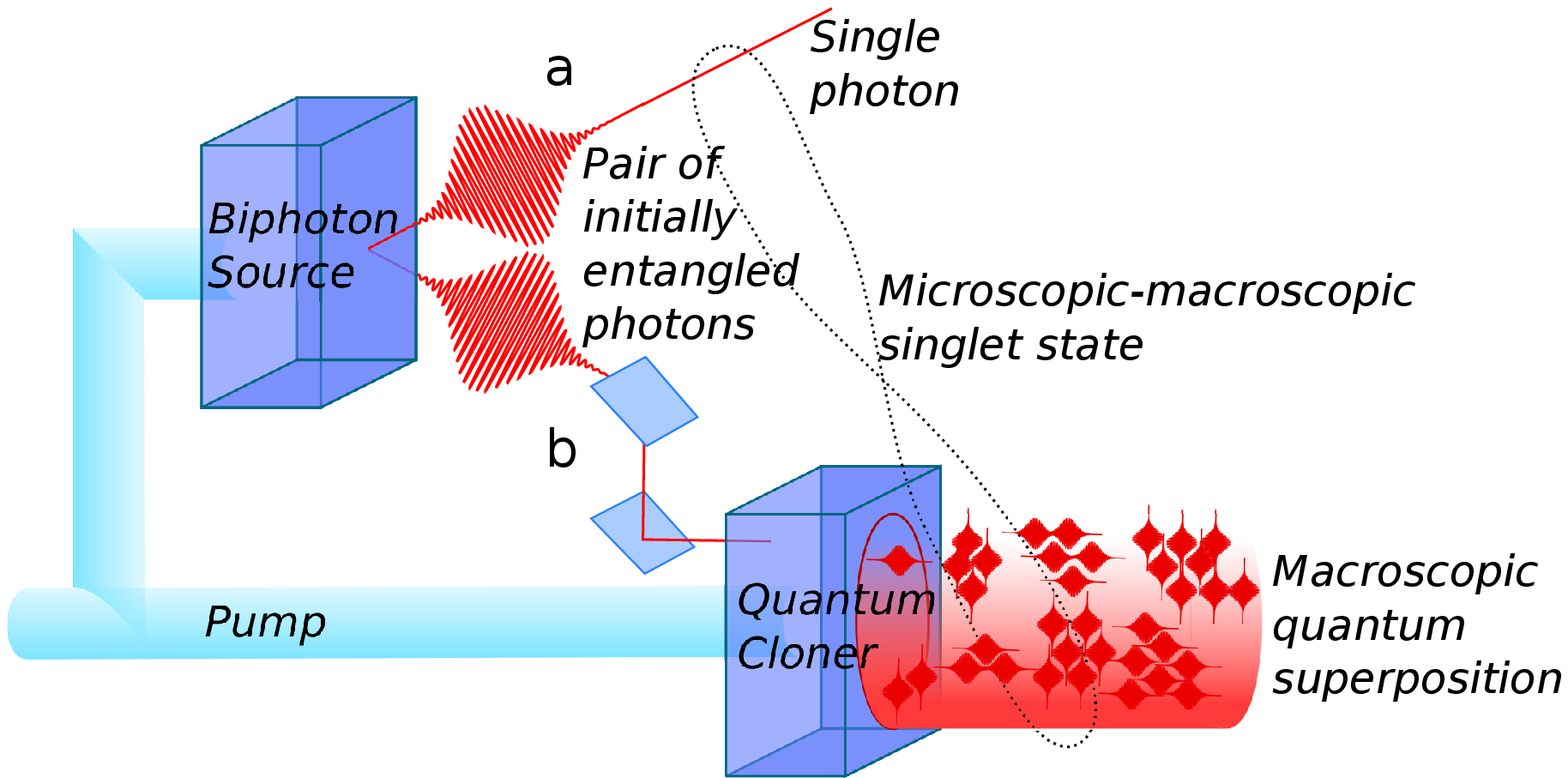}
    \quad
    \includegraphics[height=3cm]{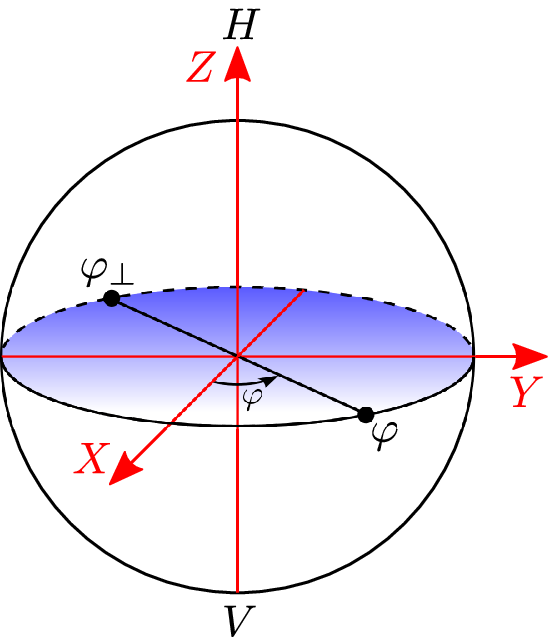}
    \raisebox{2.5cm}{b)}
  \end{center}
\caption{(Color online) a) Phase-covariant optimal quantum cloner. b) Poincar\'e polarization sphere. The blue great circle contains equatorial states of polarization.}
\label{fig:cloner}
\end{figure}

\begin{figure}
  \begin{center}
    \includegraphics[height=3cm]{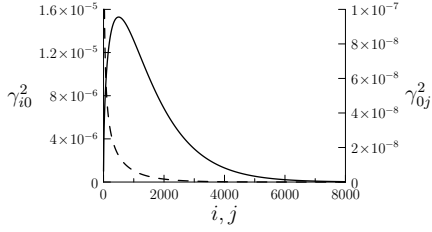}
  \end{center}
  \vskip-7mm
\caption{Probabilities $\gamma_{i0}^2$ (solid line, left Y axis) and
  $\gamma_{0j}^2$ (dashed line, right Y axis) computed for gain $g=4$.}
\label{fig:setup1}
\end{figure}

In our analysis, we focus on the micro-macro singlet state, but our results can be generalized for the bright squeezed vacuum. The singlet is produced by phase covariant optimal quantum cloner \cite{DeMartini-PRL}. This method requires first a pair of linearly polarized photons created in a usual state through parametric frequency down conversion (PDC) $1/\sqrt{2}(|1_H\rangle_a |1_V\rangle_b - |1_V\rangle_a |1_H\rangle_b)$. Subscripts $a$ and $b$ denote two spatial modes and $H$ and $V$ denote horizontal and vertical polarization, see Fig.~\ref{fig:cloner}a. The equatorial states of the Poincar\'e sphere of all polarization states, see Fig.~\ref{fig:cloner}b, are given by the following transformation of the linear polarization $a_{\varphi}^{\dagger} = 1/\sqrt{2}(e^{i\varphi} a_H^{\dagger} + e^{-i\varphi} a_V^{\dagger})$, $a_{\varphi^{\perp}}^{\dagger} = i/\sqrt{2}(e^{i\varphi} a_H^{\dagger} - e^{-i\varphi} a_V^{\dagger})$, where $a_{\varphi}^{\dagger}$ and $a_{\varphi^{\perp}}^{\dagger}$ are creators for two orthogonal polarizations $\varphi$ and $\varphi^{\perp}$. This subspace, parametrized by the polar angle $\varphi \in \langle 0,2\pi)$, is privileged for the phase covariant cloner. Here, its Hamiltonian reads $H=i \hbar \Gamma \left((a_{\varphi}^{\dagger})^2 + (a_{\varphi^{\perp}}^{\dagger})^2 \right) + \mathrm{h.c.}$ and shows that all equatorial states are cloned equally well (the form of $H$ is the same for all $\varphi$). Due to rotational invariance of the singlet, it can be expressed in this basis $1/\sqrt{2}(|1_{\varphi}\rangle_a |1_{\varphi^{\perp}}\rangle_b - |1_{\varphi^{\perp}}\rangle_a |1_{\varphi}\rangle_b)$. Next, one of its spatial modes, e.g.\ b, is amplified by the cloner to create a multi-photon state $|\Phi\rangle = e^{i H t/\hbar}|1_{\varphi}\rangle$ or $|\Phi_{\perp}\rangle = e^{i H t/\hbar}|1_{\varphi^{\perp}}\rangle$.  This unitary evolution leads to the micro-macro singlet 
\begin{eqnarray}
\lvert\Psi^-\rangle= 1/\sqrt{2}(|1_{\varphi}\rangle_a |\Phi_{\perp}\rangle_b - |1_{\varphi^{\perp}}\rangle_a |\Phi\rangle_b).
\end{eqnarray}
The multi-photon states are infinite superposition of photon number states
\begin{equation}
|\Phi\rangle = \sum_{i,j=0}^{\infty} \!\gamma_{ij}
\big|(2i+1)_{\varphi},(2j)_{\varphi^{\perp}}\rangle,\quad
|\Phi_{\perp}\rangle = \sum_{i,j=0}^{\infty} \!\gamma_{ij}
\big|(2j)_{\varphi},(2i+1)_{\varphi^{\perp}}\rangle,
\label{macro-qubits}
\end{equation}
with the real-valued probability amplitude
\begin{equation}
\gamma_{ij}=C_g^{-2}(T_g/2)^{i+j}\sqrt{(2i+1)!\,(2j)!}/(i!\,j!) = C_g^2\,\gamma_{i0}\,\gamma_{0j},
\label{gamma}
\end{equation}
where $\sum_{i,j=0}^{\infty}\gamma_{ij}^2=1$.
Here, $g = \int dt \Gamma$ is amplification gain, $C_g = \cosh (g)$, $T_g=\tanh (g)$. The notation $\big|(2j)_{\varphi},(2i+1)_{\varphi^{\perp}}\rangle$ denotes $2j$ photons in polarization $\varphi$ and $2i+1$ in $\varphi^{\perp}$, which contribute to the superposition $|\Phi_{\perp}\rangle$ with probability $\gamma_{ij}^2$. Fig.~\ref{fig:setup1} depicts $\gamma_{i0}(i)^2$ and $\gamma_{0j}(j)^2$ for $g=4$ and reveals slow decay of $\gamma_{i0}^2$ in infinity. Convergence of $\gamma_{0j}$ and $\gamma_{i0}$ to zero is slower for higher gain.

The states $|\Phi\rangle$ and $|\Phi_{\perp}\rangle$ are orthogonal due to different occupation parity in polarizations and are called macroscopic qubit. In experiment they contained $4m \simeq 10^{4}$ photons on average, where $m=\sinh^2g$. However, in this regime detection is not single photon resolving and they reveal effective overlap. Their distinguishability is efficiently quantified by photon number distributions $p_{\Phi}(k,l)=|\langle k,l|\Phi\rangle|^{2}$ and $p_{\Phi_{\perp}}(k,l)$ giving probability of having $k$ photons in polarization $\varphi$ and $l$ in $\varphi^{\perp}$ simultaneously
\begin{equation}
  p_{\Phi}(k,l) = p_{\Phi_\perp}(l,k) = \begin{cases}
    {\gamma}_{(k-1)/2,l/2}^2& \text{for odd $k$ and even $l$,}\\
    0& \text{otherwise,}
  \end{cases}
\end{equation}
where $\sum_{k,l=0}^{\infty}p_{\Phi}(k,l)=1$. Distinguishability equals
\begin{equation}
  \label{eq:overlap_symmetry}
  v= 1 - \sum_{k,l=0}^{\infty} \sqrt{p_{\Phi}(k, l)\,p_{\Phi_\perp}(k, l)}.
\end{equation}

\subsection{Additional operations performed on MQS}

Partial indistinguishability of $|\Phi\rangle$ and $|\Phi_{\perp}\rangle$ is a drawback to any quantum protocol and Bell inequality test, and needs to be fixed by special quantum state engineering. In \cite{Stobinska2011,Stobinska09,Stobinska-MDF} a filtering method was suggested to increase distinguishability of macro-qubit. It is described by the projective measurement
\begin{equation}
  \mathcal{P}^{(Th)}_{\sigma} = \sum_{k,l; k+l\ge \sigma} |k,l\rangle \langle k,l|.
  \label{theoretical-preselection}
\end{equation}
It cuts off the low photon number contributions below a threshold $\sigma$ in the initial superposition so that the preselected macro-states contain $\sigma$ photons at least distributed over two polarization modes. We will call this operation the theoretical preselection. Photon number distributions for MQS subjected to theoretical preselection read $p_{\Phi}^{(Th)}(k,l)= |\langle k,l|\mathcal{P}^{(Th)}_{\sigma}|\Phi\rangle|^2$
\begin{align}
  \label{eq:QTH1}
  p_{\Phi}^{(Th)}(k,l) = \begin{cases}
    |N_{Th}|^2{\gamma}_{(k-1)/2,l/2}^2& \text{for odd $k$, even $l$, $k+l\ge\sigma$,}\\
    0& \text{otherwise,}
  \end{cases}
 \end{align}
$p_{\Phi_\perp}^{(Th)}(k,l) = p_{\Phi}^{(Th)}(l,k)$, normalization constant $|N_{Th}|^{-2} = \sum_{i+j\ge (\sigma-1)/2}^{\infty} \gamma_{ij}^2$.

For large class of useful observables, mean values evaluated for preselected multi-photon states will involve sums of the form $|N_{Th}|^2\sum_{i,j=0,i+j\ge\sigma}^{\infty}\gamma_{ij}^2\,\tilde{f}(2i+1,2j)$, where $\tilde{f}(p,q)$ is a polynomial. For example, for mean number of photons one sets $\tilde{f}(p,q)=(p+q)$ and $\tilde{f}(p,q)=(p+q)^2$ for its variance.

In the experiment, theoretical preselection is implemented by a weak measurement. The state is split into two beams, reflected and transmitted, by a beam splitter (BS) with high transmitivity. Quantum operation $\mathcal{P}^{(Th)}_{\sigma'}$ is performed only on the reflected part and the result is feedforwarded to the transmitted beam (i.e.\ the summation constraint on occupations is shifted from reflected to transmitted part). This is the beam-splitter preselection. 

Action of a BS with reflectivity $R$ (transmitivity $T=1-R$) on two orthogonal polarizations $\varphi$ and $\varphi_{\perp}$ is independent. For a given polarization, if one BS input is vacuum $|0\rangle$ and the other is photon number state $|N\rangle$, its action is given by $\mathcal{U}_{BS}|0,N\rangle = \sum_{k=0}^N c_k^{(N)} |N-k\rangle_t |k\rangle_r$ with the probability amplitude $c_k^{(N)} = \sqrt{ \binom{N}{k} R^k\, T^{N-k} }$, where $t$ ($r$) denotes mode where $N-K$ ($k$) photons were transmitted (reflected). For the multi-photon states we get $\mathcal{U}_{BS}|\Phi\rangle = \sum_{i,j=0}^{\infty} \!\gamma_{ij} \sum_{n=0}^{2i+1} c_n^{(2i+1)} \sum_{m=0}^{2j} c_m^{(2j)} \big|(2i+1-n)_{\varphi},(2j-m)_{\varphi^{\perp}}\rangle_t |n_{\varphi},m_{\varphi^{\perp}}\rangle_r$ and similar expression for  $\mathcal{U}_{BS}|\Phi_{\perp}\rangle$. The coefficients may be grouped as follows $\gamma_{ij}\, c_n^{(2i+1)}\, c_m^{(2j)} =  C_g^2\,\gamma_{i0}\, c_n^{(2i+1)}\, \gamma_{0j}\, c_m^{(2j)}$. Evaluation of any physical quantity will lead to the following subexpressions
\begin{align}
  \label{eq:subi}
  f_i(n,i) &=C_g^2\,\gamma_{i0}^2\,\left(c^{(2i+1)}_n\right)^2
  =C_g^{-2}\left(T_g/2\right)^{2i}\!\!\frac{(2i+1)!^2}{i!^2n!(2i+1-n)!}R^nT^{2i+1-n},
\end{align}
and $f_j(m,j)=C_g^2\,\gamma_{0j}^2\,(c^{(2j)}_m)^2$, where $i,j,n,m$ are non-negative integers. For the beam-splitter preselection photon number distributions read
\begin{align}
  \label{eq:QBS1}
    p_{\Phi}^{(BS)}(k,l) =&|N_{BS}|^2\,
    \sum_{i,j=0}^{\infty}\kern0.5em
    \sum_{\substack{0\le n\le 2i+1\\ 0\le m\le 2j\\ n+m\ge \sigma'}}
    f_i(n,i)\cdot f_j(m,j)
    \cdot \delta_{k, 2i+1-n}\, \delta_{l, 2j-m},
\end{align}
with $p_{\Phi_\perp}^{(BS)}(k,l) = p_{\Phi}^{(BS)}(l,k)$, where normalization constant 
$|N_{BS}|^{-2} = \sum_{i,j= 0}^{\infty}\break\sum_{n=0}^{2i+1}\sum_{m=0,n+m\ge\sigma'}^{2j} f_i(n,i)\cdot f_j(m,j)$ and $\delta_{a,b}$ is a Kronecker delta function equal 1 if $a=b$ and 0 else. Here, important mean values are computed by $|N_{BS}|^2\sum_{i,j= 0}^{\infty}\sum_{n=0}^{2i+1}\sum_{m=0,n+m\ge\sigma'}^{2j} f_i(n,i)\cdot f_j(m,j)\,\tilde{f}(2i+1-n,2j-m)$.

Beam splitter is useful for many other basic quantum operations, e.g.\ losses, homodyne detection, inefficient detection, entanglement distillation.

\subsection{Hypergeometric function}

We notice that the ratios $(\gamma_{i+1,0}/\gamma_{i,0})^2$ and $f_i(n, i+1)/f_i(n, i)$ are rational functions of $i$ and $i$ and $n$, respectively. According to the definition of a hypergeometric term \cite{A=B}, $\gamma_{ij}^2$, $f_i$, $f_j$ are double hypergeometric terms and their infinite sums over these parameters are hypergeometric functions.

The sum of the probability governing MQS is given by Gaussian hypergeometric function
\begin{equation}
\label{eq:G}
G(n)\!=\!\!\!\sum_{i=n}^{\infty}\!\gamma_{i0}^2
= x_n\,
{}_2F_1\left(\begin{matrix}
     1, a\\
     n+1
     \end{matrix};
     z\right),\,
 \bar{G}(m)\!=\!\!\!\sum_{j=m}^{\infty}\! \gamma_{0j}^2
= y_m\,
{}_2F_1\left(\begin{matrix}
      1, b\\
      m+1
      \end{matrix};
      z\right),
\end{equation}
where $x_n = (2n+1)!\, T_g^{2n}/(C_g^2\,{n!}^2\,2^{2n})$, $y_m = (2m)!\,T_g^{2m}/(C_g^2\,{m!}^2\,2^{2m})$, $z=T_g^2$, $a = n+\frac{3}{2}$, $b = m+\frac{1}{2}$. The situation is similar for the probability governing the MQS after passing BS. Here, the sums $A(n) = \sum_{i=\lfloor n/2\rfloor}^{\infty} \!\!f_i(n,i)$ and $B(m) = \sum_{j=\lfloor (m+1)/2\rfloor}^{\infty} \!\!f_j(m,j)$, where $\lfloor x\rfloor$ is the floor function $\lfloor x\rfloor = \max \{m\in\mathbb{Z} | m\le x\}$, have to be written for odd and even $n$ and $m$ separately 
\begin{align}
  \label{eq:AB}
  A(2n) =& x_n (2n+1)\, R^{2n} T\;
  {}_2F_1\left(\begin{matrix}
      a, a\\
      \frac{3}{2}
      \end{matrix};
      T^2 z\right),
  A(2n+1) = x_n R^{2n+1}\,
  {}_2F_1\left(\begin{matrix}
      a, a\\
      \frac{1}{2}
      \end{matrix};
      T^2 z\right),\\
   B(2m) =& y_m R^{2m}\, 
   {}_2F_1\left(\begin{matrix}
      b, b\\
      \frac{1}{2}
      \end{matrix};
      T^2 z\right),
  B(2m-1) = y_m 2m\, R^{2m-1} T\, 
  {}_2F_1\left(\begin{matrix}
      b, b\\
      \frac{3}{2}
      \end{matrix};
      T^2 z\right). \nonumber     
\end{align}
We checked that no closed form of $G(n)$, $\bar{G}(m)$, $A(n)$ and $B(m)$ exist using the Gosper's, Zeilberger's, and Petkov\v{s}ek's algorithms \cite{A=B}. These algorithms constitute the standard mathematical tools for hypergeometric series analysis.  Closed form, if existed, would eliminate sums over $i$ and $j$ completely and reduce the computation time.  

In our case, these sums have to be computed iteratively. This prevents efficient simulation of MQS properties and evolution for realistic parameters, since it is neither possible to bring this function to elementary ones nor to other special functions (e.g.\ Bessels) tractable more easily. The function ${}_2F_1$  arises directly form probability governing the MQS $\gamma_{ij}^2$. Parameters $a$ and $b$ of ${}_2F_1$ are half integers.  This class of Gaussian hypergeometric functions is the least known in the literature.  There are only a few known identities simplifying ${}_2F_1$ with half-integer parameters \cite{A=B,Abramowitz}, but they cannot be used here. Moreover, recurrence transformations lead to other half-integer forms of ${}_2F_1$. Unconditional convergence of the hypergeometric sum is guaranteed by $z,T^2\in(0,1)$.

It is difficult to find the appropriate cutoffs for the sums over $i$ and $j$. High cutoffs make the computation long and error prone whereas too low values result in omitting significant terms. The fast computation of ${}_2F_1$ in a non-asymptotic regime of its arguments and parameters is challenging \cite{Michel2008,Lopez2010}. In literature, there are algorithms applicable for some special cases of relations between its parameters \cite{Abramowitz,Temme2003,Miller2003}, but not useful here. Moreover, in our intermediate regime they become unstable. Neither Gaussian quadrature \cite{Gautschi2002} nor Pad\'e \cite{Luke1972} approximations exist.

To be able to compute Eq.~(\ref{eq:QBS1}) and similar ones of the form
\begin{equation}
  \label{eq:sumij}
  S = \sum_{i,j=0}^{\infty} \sum_{n=0}^{2i+1} \sum_{m=0;n+m \ge {\sigma'}}^{2j}
      f_i(n,i)\, f_j(m,j)
\end{equation}
we aim at partial factorization with respect to $(i,n)$ and $(j,m)$ to compute the sums over $i$ and $j$ separately. Full factorization cannot be achieved due to the preselection condition $n+m\ge {\sigma'}$.  We change the variable order $\sum_{i=0}^{\infty}\sum_{n=0}^{2i+1} f_i(n,i) = \sum_{n=0}^{\infty}\sum_{i=\lfloor n/2\rfloor}^{\infty} f_i(n,i)$ (similarly for $f_j(m,j)$) in Eq.~(\ref{eq:sumij})
\begin{equation}
  \label{eq:sumij_rewritten}
  S = \sum_{n,m=0; n+m\ge {\sigma'}}^{\infty} A(n)\, B(m).
\end{equation}
We avoid four nested sums and get a double sum of independent series. However, it became clear, that the hypergeometric functions have to be additionally summed up in infinite sums over its parameters.

In case of computation of mean values, $\tilde{f}(p,q)$ has first to be split into sum of monomials $\tilde{f}(p,q)=\sum_{n,m} a_{nm}\,p^n\,q^m$, where, importantly, $a_{nm}$ are real coefficients. Then hypergeometric terms take the form $\tilde{f}_i(n,i)=f_i(n,i)\cdot (2i+1-n)^p$ and $\tilde{f}_j(m,j)=f_j(m,j)\cdot (2j-m)^q$, where $p$ and $q$ are integers. The analysis presented above for $f_i(n,j)$ still holds for $\tilde{f}_i$ and $\tilde{f}_j$.

\section{Numerical Methods}
\label{sec:numerical_methods}

\subsection{Calculation of Hypergeometric Terms with High Precision}
\label{sssec:subexpr}

Factorials in numerators and denumerators of hypergeometric terms $\gamma_{i0}$, $\gamma_{0j}$, $f_i$ and $f_j$ take large values, unavailable for the standard machine arithmetic.  IEEE 754 double precision numbers are capable of storing 15 significant decimal digits and magnitudes $10^{-308}$ to $10^{308}$ \cite{ieee754}.  For $i,j \ge 50$ they exceed $10^{308}$, but the values of terms are between 0 and 1 and need to be computed precisely.

One way to compute $\gamma_{i0}^2$, $\gamma_{0j}^2$, $f_i$ and $f_j$ is to use libraries offering enhanced precision arithmetic. For example, the GNU Multiple Precision Arithmetic Library (GMP) \cite{gmp} and the GNU Multiple Precision Floating-Point Reliable Library (MPFR) \cite{mpfr} are available for the C and C++. The Class Library for Numbers (CLN) \cite{cln} cooperates with C++ and the mpmath \cite{mpmath} with Python. These libraries offer low speed compared to the double precision arithmetic.

The other possible way of computing hypergeometric terms is to use their logarithmic forms: $\log\gamma_{i0}$, $\log\gamma_{0j}$, $\log f_i$, $\log f_j$. The number magnitudes are within range of double precision for large $i$ and $j$.  Moreover, both the standard C/C++ and MPFR offer \verb|lgamma(n)| function, which calculates $\log(\Gamma(n))=\log((n-1)!)$ with hardware acceleration.

In order to find the best method of computation of the subexpressions we compared all the mentioned solutions.  The relative error estimation was based on the values obtained symbolically with computer algebra system Wolfram Mathematica. We showed superiority of C/C++ computation with double-precision.  It is not only several orders of magnitude faster but also consumes less memory. The reason for that is that IEEE 754 arithmetic is built-in into modern computer processors and does not need any additional memory structures.  The average error of double-precision calculation is of the order of $10^{-12}$, low enough for numerical simulations. The error can be further decreased at the significant cost of speed with MPFR library, which is the second fastest solution.

\subsection{Convergence Rate}
\label{sssec:convergence}

Computation of $G(n)$, $\bar{G}(m)$, $A(n)$ and $B(m)$ requires finding proper cutoffs for infinite sums over $i$ and $j$ in Eqs.~(\ref{eq:G}) and (\ref{eq:AB}). First, we tested convergences of these sums, since their acceleration allows setting cutoffs earlier. Several convergence rate acceleration methods were tested: the Aitken's delta-squared process \cite{aitken,nr}, the Shanks transformation \cite{shanks} and the Richardson extrapolation \cite{richardson,brezinski}. None of them improves much the computation time. They raise the complexity of the algorithm instead.

\subsection{Efficient Computation of the Gaussian Hypergeometric Function}
\label{sssec:cutoff}

\begin{figure}[t]
  \begin{center}
    \includegraphics[height=3cm]{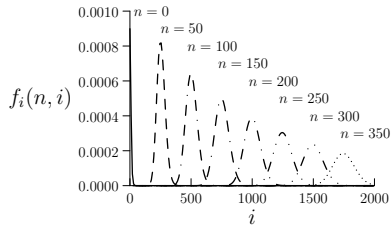}
  \end{center}
  \vskip-7mm
  \caption{Function $f_i(n,i)$ evaluated for $i$, parametrized by $n$ ($g=4$).}
  \label{fig:fij}
\end{figure}

Next, we worked out algorithm for finding cutoffs for sums $A(n)$ and $B(m)$, but simplified procedure holds for $G(n)$ and $\bar{G}(m)$.  The hypergeometric terms $\gamma_{i0}^2$, $\gamma_{0j}^2$, $f_i(n,i)$ and $f_j(m, j)$ converge slowly. Function $f_i$ is depicted in Fig.~\ref{fig:fij} (see Fig.~\ref{fig:setup1} for $\gamma_{ij}$).  The greater $n$ gets, the slower $f_i$ converge over variable $i$ and more terms have to be included in the summation (similarly for $f_j$ and variable $j$). One may apply high constant sum cutoffs. But even for large $i$ ($j$) there exist $n$ ($m$) for which the fixed range is too small.

As a solution to this problem, we applied a dynamic cutoff with an adjustable precision e.g.\ $10^{-15}$. The summation range was divided into two intervals.  In the first one, the terms $f_i$ and $f_j$ monotonically grow to achieve the global maximum. The second interval is infinite and the terms slowly decrease asymptotically to 0. The value $i=I$ for which $f_i$ achieves its maximum can be found by solving differential equation $\frac{\partial f_i(n,i)}{\partial i} = 0$, which leads to equation $\Psi_0(2I+1-n+1) = \Psi_0(n+1) + \log R - \log T$, where $\Psi_0(x)$ denotes the digamma function (the first derivative of $\log\Gamma(x)$).  The inverse of $\Psi_0(2I+1-n+1)$ can be computed numerically with the Paul Fackler's method \cite{fackler}. Similar method may be used for finding maximum of $f_j$.

Alternatively, algorithm starts from $i=\lfloor n/2\rfloor$ ($j=\lfloor (m+1)/2\rfloor$) and for each $i$ ($j$) evaluates $f_i(n,i)$ ($f_j(m,j)$). It adds this term to the sum and compares $f_i(n,i)$ to the previous term $f_i(n,i-1)$.  If $f_i(n,i) > f_i(n,i-1)$ the algorithm is still in the first interval and continues summation. Otherwise, the summation is performed over the second interval, the algorithm checks if $f_i$ is smaller than the desired precision and if yes, it stops. Error is minimized by adding the terms in the increasing order in the first interval. In practice, the algorithm uses the logarithmic forms (they allow finer comparison of $f_i(n,i)$ and $f_i(n,i-1)$).

We compared the fixed and dynamic methods of assigning cutoffs for the sum $A(n)$ in Table~\ref{tab:fijcutoff} (the numbers for $B(m)$ are similar).  
The latter decreases the number of terms for small $n$ and $m$ and achieves a good precision for their large values. It significantly improves speed and accuracy of the computation e.g.\ for $g=4$, if $n=100$, the summation includes $1000$ whereas if $n=0$ it includes $150$ terms to get the same accuracy with relative error $10^{-15}$.

\begin{table}
  \begin{center}
    \small
    \begin{tabular}{crrcrc}\toprule[1pt]
      \textbf{Sum}& \multicolumn{3}{c}{\textbf{Fixed sum cutoff}}&
      \multicolumn{2}{c}{\textbf{Dynamic sum cutoff}}\\
      & \multicolumn{1}{c}{{Cutoff}}&
      \multicolumn{1}{c}{{Value}}&
      \multicolumn{1}{c}{{Relative}}&
      \multicolumn{1}{c}{{Value}}&
      \multicolumn{1}{c}{{Relative}}\\
      & & & \multicolumn{1}{c}{{error}}&
      & \multicolumn{1}{c}{{error}}
      \\\midrule[1pt]
      $A(0)$
      & 50& $0.01079$& $10^{-6}$&
      $0.01079$& $10^{-15}$\\
      & 100& $0.01079$& $10^{-9}$\\
      & 200& $0.01079$& $10^{-15}$\\
      & 500& $0.01079$& $10^{-15}$\\\midrule[0.5pt]
      $A(100)$
      & 500& $0.03695$& $10^{-1}$&
      $0.07634$& $10^{-15}$\\
      & 750& $0.07634$& $10^{-6}$\\
      & 1000& $0.07634$& $10^{-15}$\\
      & 2000& $0.07634$& $10^{-15}$\\\midrule[0.5pt]
      $A(200)$
      & 1000& $0.03286$& $10^{-1}$&
      $0.06546$& $10^{-15}$\\
      & 1500& $0.06546$& $10^{-10}$\\
      & 2000& $0.06546$& $10^{-15}$\\
      & 2500& $0.06546$& $10^{-15}$\\
      \bottomrule[1pt]
    \end{tabular}
  \end{center}
  \caption{Comparison of cutoff values and relative error of computation of $A(n)$ for $g=4$ with the fixed and dynamic cutoff estimation methods.}
  \label{tab:fijcutoff}
\end{table}

$\tilde{f}_i$ and $\tilde{f}_j$ have the same properties as $f_i$ and $f_j$, only converge slower.  Therefore, the presented algorithm is capable of finding cutoffs of infinite hypergeometric sums required in computation of mean values.

\subsection{Summation of Hypergeometric Functions}

Computation of the sum $S$ in Eq.~(\ref{eq:sumij_rewritten}) involves infinite summation of product of two Gaussian hypergeometric functions $A(n)$ and $B(m)$ over $n$ and $m$, where $n+m\ge\sigma'$.  The solution is to pre-compute the values of $A(n)$ and $B(m)$ with method presented in Section \ref{sssec:cutoff} and store them in the computer memory. The next step is to perform the final summation.  The cutoffs $N$ and $M$ over variables $n$ and $m$, respectively, are found separately during the pre-computation stage, so that $\lvert A(N)/A(\sigma')\rvert$ and $\lvert B(M)/B(\sigma')\rvert$ are smaller than the desired relative computation error. 

Additional speedup of computations is achieved by changing the order of summation and eliminating one of the infinite sums
\begin{align}
  \label{eq:S1}
  S=&\sum_{n={\sigma'}}^{\infty}\,\sum_{m=0}^n A(n-m)\cdot B(m).
\end{align}
The computation time of this formula is polynomial $O(N^2)$.

Further optimization is possible by noting the fact that hypergeometric terms $\gamma_{ij}^2$, $f_i$, $f_j$ are probability distributions and $\sum_{i,j=0}^{\infty}\gamma_{ij}^2=1$, $\sum_{i,j=0}^{\infty}\sum_{n=0}^{2i+1}\break\sum_{m=0}^{2j}f_i(n,i)\,f_j(m,j)=1$, which implies $\sum_{n,m=0}^{\infty}A(n)\,B(m)=1$. These properties allow to rewrite Eq.~(\ref{eq:S1}) into a form involving finite summation range
\begin{align}
  \label{eq:S2}
  S=&1- \sum_{n=0}^{{\sigma'}} A(n)\sum_{m=0}^{{\sigma'}-n} B(m).
\end{align}
Here computation time is also polynomial $O({\sigma'}^2)$.

Eq.~(\ref{eq:S2}) takes advantage over Eq.~(\ref{eq:S1}) for small values of $\sigma'$, for which the sum intervals are relatively short and the result is obtained quickly.  However, the larger value of $\sigma'$ is, the more terms in Eq.~(\ref{eq:S2}) have to be taken into account.  As a result of that, the summation takes longer and errors accumulate significantly.  At some point Eq.~(\ref{eq:S1}) gives more reliable results and performs faster than Eq.~(\ref{eq:S2}).  Thus, the routine computing $S$ decides, depending on $\sigma'$, which method to apply.

\section{Examples and Applications}
\label{sec:examples}

\subsection{Computation of MQS Normalization and Photon Number Distribution}
\label{ssec:qfunction}

Formulae for photon number distributions, e.g.\ Eq.~(\ref{eq:QTH1}) or (\ref{eq:QBS1}) require prior computation of normalization $|N_{Th}|^2$ and $|N_{BS}|^2$, respectively. In principle, it is possible to take advantage of the fact that $\sum_{k,l=0}^{\infty} p_{\Phi}(k,l)=1$, compute unnormalized distribution values for all pairs $(k,l)$, sum them up to obtain the inverse of its normalization constant and later renormalize the result. This method is useful in cases where all meaningful values of $p_{\Phi}(k,l)$ have to be computed anyway. However, in general, the normalization constant should be computed separately. 

For theoretical preselection, according to Eqs.~(\ref{eq:S1}) and (\ref{eq:S2}), normalization constants equal
\begin{align}
  \label{eq:NTH1b}
  \frac{1}{|N_{Th}|^2} &= C_g^{4}\kern-1em
  \sum_{i=(\sigma-1)/2}^{\infty}\kern.25em
  \sum_{j=0}^{i}
  \gamma_{(i-j),0}^2\cdot
  \gamma_{0j}^2,
  \\
  \frac{1}{|N_{Th}|^2} &= 1 - C_g^{4} \sum_{i=0}^{(\sigma-1)/2} \gamma_{i0}^2
  \sum_{j=0}^{(\sigma-1)/2-i} \gamma_{0j}^2.
   \label{eq:NTH1c}
\end{align}
We empirically found for $g=4$, $\sigma\approx 5000$ to be a good threshold for switching from Eq.~(\ref{eq:NTH1c}) to Eq.~(\ref{eq:NTH1b}) . According to Eq.~(\ref{eq:QTH1}), computation of $|N_{Th}|^2$ directly gives $p_{\Phi}^{(Th)}(k,l)$.

$|N_{BS}|^2$ is computed directly with Eqs.~(\ref{eq:S1}) and (\ref{eq:S2}).
We found that for $g=4$ and ${\sigma'}\approx 500$ the computation time required by Eqs.~(\ref{eq:S1}) and (\ref{eq:S2}) is similar and for ${\sigma'}> 500$ the former is faster. In order to compute photon number distribution $p_{\Phi}^{(BS)}(k,l)$ we turn Eq.~(\ref{eq:QBS1}) into a simpler form 
\begin{equation}
  \label{eq:QBS1c}
    p_{\Phi}^{(BS)}(k,l) =|N_{BS}|^2\,
    \sum_{n=\sigma'}^{\infty}\,\sum_{m=0}^n
    \begin{cases}
      \begin{aligned}
        &f_i\left(\tfrac{k+n-m-1}{2},n-m\right)\\
        &\quad \cdot f_j\left(\tfrac{l+m}{2},m\right)
      \end{aligned}
      &
      \begin{aligned}
        &\text{if $k+n-m$ is odd}\\
        &\text{\quad and $l+m$ is even,}
      \end{aligned}
      \\
      0& \text{otherwise.}
    \end{cases}
\end{equation}
It is similar to Eq.~(\ref{eq:NTH1b}) and is evaluated likewise. The main difference is that sums in Eq.~(\ref{eq:QBS1c}) traverse only even or odd values of $n$ and $m$, depending on the parity of $k$ and $l$.  The computation time is polynomial $O(N^2)$, where $N$ is the cutoff of the infinite sum over $n$.

\subsection{Computation of MQS Distinguishability}
\label{ssec:overlap}

Distinguishability given in Eq.~(\ref{eq:overlap_symmetry}) is not optimal for computation, since it does not minimize summation errors and for large values of the cutoffs it is difficult to keep coefficients of the photon number distributions in memory.

\subsubsection{Optimization of Mathematical Formulae}
\label{sssec:optimization}

Eq.~(\ref{eq:overlap_symmetry}) may be optimized by change of variables 
\begin{equation}
  \label{eq:overlap2}
  v=1 - 2\sum_{k=0}^{K}\,\sum_{l=0}^{k-1} \sqrt{p_{\Phi}(k, l)\,p_{\Phi}(l, k)}
  - \sum_{k=0}^{K} p_{\Phi}(k, k),
\end{equation}
where $K$ is the cutoff of the infinite sum over variable $k$.  The
computation time of Eq.~(\ref{eq:overlap2}) is polynomial $O(K^2)$ and
has the same advantages and disadvantages as
Eq.~(\ref{eq:overlap_symmetry}), but requires only half of the
calculations.

More complex idea is to first, divide the $(k,l)$ plane into rectangles, compute the photon number distributions for each rectangle separately and save the data to a disk. Next, we calculate partial overlaps for each rectangle and sum up the contributions \cite{chhajlany}. Our approach uses tiles of a size $K'\times L'$ and performs two-step summation
\begin{equation}
  \label{eq:overlap3}
    v \!=\! 1-\sum_{k=0}^{K/K'}\,\sum_{l=0}^{L/L'}\Biggl(
      \sum_{k'=0}^{K'-1}\,\sum_{l'=0}^{L'-1}
      \sqrt{p_{\Phi}(x, y)\,p_{\Phi}(y, x)}\Biggr),
\end{equation}
where $x=K'k+k'$, $y=L'l+l'$ and $K$, $L$ are the cutoffs of the sums over $k$ and $l$ respectively. Computation complexity of Eq.~(\ref{eq:overlap3}) is the same as of Eq.~(\ref{eq:overlap_symmetry}), but this method offers several advantages.  The values of $K'$ and $L'$ can be selected in such a way that the precomputed values of coefficients (e.g.\ $A(n)$ and $B(m)$) are stored in memory and used for all points of the single tile. Additionally, different tiles can be computed in parallel with separate processors. This significantly speeds up the calculations and leads to a better usage of the computer resources and minimizes errors by grouping summed terms in smaller sets, but requires proper estimation of cutoffs $K$, $L$ and tile size $K'\times L'$. Computation gets twice shorter if Eq.~(\ref{eq:overlap3}) is rewritten to the following form
\begin{equation}
  \label{eq:overlap3a}
  \begin{split}
    v=&1-2\sum_{k=0}^{K/K'}\,\sum_{l=0}^{k-1}\Biggl(
      \sum_{k'=0}^{K'-1}\,\sum_{l'=0}^{K'-1}
      \sqrt{p_{\Phi}(x, w)\,p_{\Phi}(w, x)}\Biggr)\\
      &-\sum_{k=0}^{K/K'}\Biggl(\sum_{k'=0}^{K'-1} p_{\Phi}(x, x)
      +2\sum_{l'=0}^{k'-1}\sqrt{p_{\Phi}(x, w)\,p_{\Phi}(w, x)}
      \Biggr),
  \end{split}
\end{equation}
where $w=K'l+l'$.

\subsubsection{Approximation: Cutoffs of Infinite Sums}
\label{sssec:approximation_cutoffs}

\begin{figure}
  \begin{center}
    \includegraphics[height=2cm]{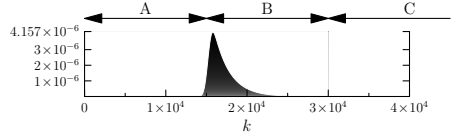}
  \end{center}
  \vskip-7mm
  \caption{The plot of $p_{\Phi}^{(BS)}(k,0)$ showing three main intervals of $k$ where
    photon number distribution behaves differently.}
  \label{fig:ranges}
\end{figure}

In order to estimate cutoffs $K$ and $L$ in Eqs.~(\ref{eq:overlap3}) and (\ref{eq:overlap3a}) we tried the method presented in Section~\ref{sssec:cutoff}. However, unlike $A(n)$ and $B(m)$, Eq.~(\ref{eq:overlap3}) involves a lot of near-zero terms for small $k$ and $l$ before entering the more significant regions and we had to adjust this algorithm accordingly.  The new solution analyzes the shape of photon number distributions. The distribution cut for a given $l$ spans of three different intervals of $k$, see Fig.~\ref{fig:ranges}.
The first interval (A) starts at $k=0$ and ends at $k\approx 10\,{\sigma'}$ for the beam-splitter preselection ($10\,\sigma$ for the theoretical). Here, the values of the distributions are small and their input is negligible.  The interval B for $k\in\langle 10\,{\sigma'}, 30m\rangle$ ($k\in\langle 10\,\sigma, 20m\rangle$) gives the main input. In region C the values of are slowly fading out in infinity. 

The proposed algorithm analyzes the values of $p_{\Phi}(k,0)$ to find the start of the region C\@.  To speed up the search, it begins at $k=30m$ ($k=20m$), which is an approximate end of region B, and takes samples of the photon number distribution for
$k=30m,30m+100,30m+200\ldots$ to the point where the values of the distribution are not distinguishable from 0 with the desired precision.  This point lays inside region C and can be used as a good cutoff value $K$ and $L$.  This method gives the results which confirm the values obtained analytically.

\subsection{Simulation of Finite Detector Resolution}

For the theoretical preselection we model the resolution of detectors measuring photon number $\pm150$ with the Weierstrass transform, i.e.\ we apply the low-pass filter with Gaussian distribution for which $3\bar{\sigma}\approx 150$.  The photon number distribution ${p'}_{\Phi}^{(Th)}(k,l)$ takes the form
\begin{equation}
  \label{eq:weierstrass}
  {p'}_{\Phi}^{(Th)}(k,l) =
  \frac{1}{2\pi\bar{\sigma}^2}
  \sum_{p,q=-3\bar{\sigma}}^{3\bar{\sigma}}
  p_{\Phi}^{(Th)}(k-p, l-q)\,e^{\frac{-(p^2+q^2)}{2\bar{\sigma}^2}},
\end{equation}
where $\bar{\sigma}$ is the standard deviation. Gaussian properties of the filter allow for replacing the infinite summation over $p$ and $q$
with finite ranges.

In case of the beam-splitter preselection, applying a Gaussian filter to the photon number distribution values does not influence the result in a significant way.  Since the filtering slows the computations, which in case of the beam-splitter preselection are already time-consuming, there is no reason for the Weierstrass transform.

Application of Eq.~(\ref{eq:weierstrass}) to Eq.~(\ref{eq:overlap3a}) means that each tile of size $K'\times K'$ has to be increased by a $3\bar{\sigma}$ margin.  Computation time for each tile is increased approximately $(1+6\bar{\sigma}/K')^2$ times and requires more computer memory, but is still quite easily tractable.

\section{MQSVIS Program Suite}

MQSVIS is a program suite developed for computation of various quantum
indicators for MQS of light.  These indicators include:
\begin{itemize}
\item photon number distributions for MQS of light,
\item visibility,
\item visibility computed from the overlap,
\item visibility computed for the photon number distributions
  transformed with Weierstrass transform for different
  values of $3\sigma \in\{1, 1.5, 15, 150\}$  (Gaussian blur),
\item mean number of photons in both polarizations (jointly and separately
  for each polarization),
\item variance of number of photons in both polarizations (jointly and
  separately for each polarization).
\end{itemize}

\subsection{Compilation of the Programs}

In order to compile and test MQSVIS program suite for Linux/Unix
operating systems with GNU utilities (GNU make, GNU compiler
collection) it is enough to run the following commands in a directory
containing unpacked source code
\begin{verbatim}
    make
    make check
\end{verbatim}
In case of Linux/Unix with GNU make and Intel C Compiler, one has to
modify \verb|Makefile|, replacing \verb|gcc| with \verb|icc| and
\verb|-fopenmp| option with \verb|-openmp|.

For other platforms and compilers: one should compile
\verb|mqsvis_norm.c| with a standard C compiler, and
\verb|mqsvis_tile.c| with a C compiler and OpenMP extensions turned
on.  If compiler does not offer OpenMP extensions, the program will
still work but will not utilize multiple cores or processors.

\subsection{Description of the Program Suite}

\subsubsection{MQSVIS\_norm}

Program computes squared normalization for preselected macroscopic
quantum superpositions (MQS) of light. This value is required by
\emph{MQSVIS\_tile} program.

Program parameters: \emph{m} -- average number of photons, \emph{Dth}
-- preselection threshold.

\subsubsection{MQSVIS\_tile}

Program computes photon number distribution for preselected
macroscopic quantum superpositions (MQS) of light for a single tile.
It produces partial results of quantum indicators: visibility, man number
of photons, variance, to be gathered by \emph{MQSVIS\_gather} script.
Additionally, saves photon number distribution for single MQS state to
a file.

Program parameters:
\emph{m}          -- average number of photons,
\emph{Dth}        -- preselection threshold,
\emph{R}          -- amount of losses,
\emph{N2}         -- squared normalization, computed by \emph{MQSVIS\_norm} program,
\emph{tilesize}   -- width and height of a single tile,
\emph{tilex}      -- tile column (counted from 0),
\emph{tiley}      -- tile row (counted from 0),
\emph{plotstep}   -- step used for saving photon number distributions
\emph{plotfname1} -- file to save photon number distribution for tile $(\text{tilex}, \text{tiley})$,
\emph{plotfname2} -- file to save photon number distribution for tile $(\text{tiley}, \text{tilex})$.

Environment variables (used only when compiled with OpenMP extensions):
\emph{OMP\_NUM\_THREADS} -- maximal number of cores or processors utilized by the program.

\subsubsection{MQSVIS\_gather.py (Python script)}

A script developed to gather partial results computed by
\emph{MQSVIS\_tile} program.  It computes final quantum indicators for
macroscopic quantum superpositions of light.

Program requires partial results of \emph{MQSVIS\_tile} program to be saved in
text files with the names in the form
\begin{verbatim}
    M%s_Dth%s_r%s-%d,%d.txt
\end{verbatim}
where first \verb|%s| is replaced with average number of photons,
second \verb|%s| with preselection threshold, third \verb|%s| with losses and
\verb|%d,%d| are comma-separated tile coordinates numbered from 0.

Program parameters:
\emph{m}     -- average number of photons,
\emph{Dth}   -- preselection threshold,
\emph{R}     -- losses,
\emph{tiles} -- number of tiles in a row/column.

\subsection{Examples and Testing}

Below we present a sample session with the above programs.  The
sequence of commands could be used e.g.\ for testing of the suite.

\begin{enumerate}
\item Compute normalization for $m = 5$ and $Dth = 2$ and save it to a file
\begin{verbatim}
    mqsvis_norm 5 2 > normalization.txt
\end{verbatim}

\item Compute tile $(0, 0)$ for $R = 0$ (no losses), tile size $10 \times 10$
\begin{verbatim}
    mqsvis_tile 5 2 0 $(< normalization.txt) 10 0 0 1 \
        plot-0,0.txt /dev/null > M5_Dth2_r0-0,0.txt
\end{verbatim}

\item Compute tile $(1, 0)$, and parallely tile $(0, 1)$ for the same parameters
\begin{verbatim}
    mqsvis_tile 5 2 0 $(< normalization.txt) 10 1 0 1 \
        plot-1,0.txt plot-0,1.txt > M5_Dth2_r0-1,0.txt
\end{verbatim}

\item Compute tile $(1, 1)$
\begin{verbatim}
    mqsvis_tile 5 2 0 $(< normalization.txt) 10 1 1 1 \
        plot-1,1.txt /dev/null > M5_Dth2_r0-1,1.txt
\end{verbatim}

\item Gather the results
\begin{verbatim}
    python mqsvis_gather.py 5 2 0 2
\end{verbatim}
\end{enumerate}

The computed results for the above session, printed out in the last
step, are as follows
\begin{verbatim}
    total probability sum=0.626948016783147
    visibility=0.683
    visibility prim=0.683
    simple visibility computed from the overlap=1.000
    visibility with Gaussian blur (3sigma=1)=0.968
    visibility with Gaussian blur (3sigma=1.5)=0.746
    visibility with Gaussian blur (3sigma=15)=0.491
    visibility with Gaussian blur (3sigma=150)=0.912
    mean=8.362
    mean k=6.269
    mean l=2.092
    variance=68.303
    variance k=40.241
    variance l=15.715
    maximal value=0.0408525598455265
\end{verbatim}

\section{Conclusions}

We discussed numerical methods useful for modeling of macroscopic quantum superpositions of light generated by phase covariant quantum cloner. Properties of these MQS are governed by Gaussian hypergeometric function, which parameters are half integers and argument takes moderate values. No simplifications of this function are possible. It is given by slowly convergent infinite sum, impossible to accelerate. The algorithm dynamically finds cutoff as well as cutoffs of sums involving hypergeometric functions. We achieved precision~$10^{-15}$.

The model allows simulation of experimental schemes involving linear optical elements and MQS, as well as computation of expectation values of some observables.  It is not possible to obtain such results by analytical calculation. Our model takes into account values of experimental parameters. Depending on them optimizes computation by choice of the more efficient, parallelizable algorithm for evaluation of the indicators for these states, e.g.\ the photon number distribution or distinguishability. Numerical methods presented here can be adjusted for analysis of decoherence and similar experimental schemes, different filtering techniques and modified MQSs. As an example, we included a model of a realistic detector what improved applicability of our model for scientific research. We verified our numerical results with recently developed analytical model for MQS, available only for certain regimes of parameter values \cite{Stobinska2011}.

Finally, MQSVIS program suite was developed and implemented in C and
Python programming languages. It utilizes the developed numerical
model and allows to compute various quantum indicators for MQS of
light: photon number distributions, visibility in case of limited
resolution of photodetectors, mean and variance of photon number
(jointly and separately for each polarization).  Parallelization was
achieved with splitting the computing tasks into tiles and utilization
of OpenMP compiler extensions.

\section*{Acknowledgments}

A. B. thanks J. Arabas and R. W. Chhajlany for discussions. Calculations were carried out at CI TASK (Galera cluster) and Cyfronet (Zeus cluster).  This work was partially supported by Ministry of Science and Higher Education Grant No.\@ 2619/B/H03/2010/38.

\end{document}